\documentclass[amsmath,amssymb,journal=cgd, manuscript=article]{achemso}
\usepackage{graphicx}
\usepackage{hyperref}
\mciteErrorOnUnknownfalse
\hypersetup{pdfencoding=auto,colorlinks=true,linkcolor=blue,citecolor=blue}

\title{Scale-dependent optimized homoepitaxy of InAs(111)A}

\author{Steffen Zelzer}
\affiliation{Center for Quantum Devices, Niels Bohr Institute, University of Copenhagen, 2100 Copenhagen, Denmark}

\author{Rajib Batabyal}
\affiliation{Center for Quantum Devices, Niels Bohr Institute, University of Copenhagen, 2100 Copenhagen, Denmark}

\author{Derek Dardzinski}
\affiliation{Carnegie Mellon University, Pittsburgh, PA 15213, United States}

\author{Noa Marom}
 \affiliation{Department of Materials Science and Engineering, Carnegie Mellon University, Pittsburgh, PA 15213, USA}
 \affiliation{Department of Physics, Carnegie Mellon University, Pittsburgh, PA 15213, USA}
 \affiliation{Department of Chemistry, Carnegie Mellon University, Pittsburgh, PA 15213, USA}

\author{Kasper Grove-Rasmussen}
\email{k\_grove@nbi.ku.dk}
\affiliation{Center for Quantum Devices, Niels Bohr Institute, University of Copenhagen, 2100 Copenhagen, Denmark}

\author{Peter Krogstrup}
\affiliation{Center for Quantum Devices, Niels Bohr Institute, University of Copenhagen, 2100 Copenhagen, Denmark}

\date{\today}

\begin{document}
\maketitle

\begin{abstract}
    We combined in-situ scanning tunneling microscopy (STM) with the conventional growth characterization methods of atomic force microscopy (AFM) and reflection high energy electron diffraction (RHEED) to simultaneously assess atomic scale impurities and the larger scale surface morphology of molecular beam epitaxy (MBE) grown homoepitaxial InAs(111)A.
    By keeping a constant substrate temperature and indium flux while increasing the As$_2$ flux, we find two differing MBE growth parameter regions for optimized surface roughness on the macro- and atomic scale. In particular, we show that a pure step-flow regime with strong suppression of hillock formation can be achieved, even on substrates without intentional offcut. On the other hand, an indium adatom deficient, low atomic defect surface can be observed for a high hillock density. We identify the main remaining point defect on the latter surface by comparison to STM simulations. Furthermore, we provide a method for extracting root-mean-square surface roughness values and discuss their use for surface quality optimization by comparison to scale dependent, technologically relevant surface metrics. Finally, mapping the separately optimized regions of the growth parameter space should provide a guide for future device engineering involving epitaxial InAs(111)A growth.
\end{abstract}

\begin{figure}[!ht]
    \includegraphics[width=0.5\linewidth]{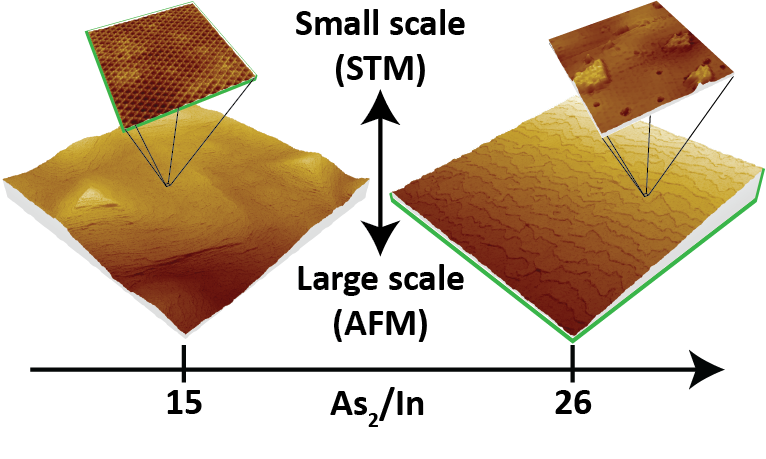}
\end{figure}

\section{Introduction}
    MBE growth of (001) oriented substrates is well studied in literature, while the (110) and (111)A, B orientations of InAs are examined in much less detail. The main reason being the formation of so-called hillocks, faceted surface protrusions of varying shape appearing for wide ranges of growth conditions. Formation of these is often attributed to lower arsenic sticking coefficients to the latter three crystal orientations, as stated in a recent review \cite{Yerino2017}. Using few degrees off-cut substrates has repeatedly been shown to help in reducing hillocks and opening up the growth windows, however, often resulting in large atomic steps, so-called "macrosteps".\cite{Yerino2017}
    
    Recently, the interest for (111) oriented substrates has increased due to a manifold of benefits over the (001) standard. (111)-oriented III-V compound semiconductors were proposed as growth templates for high quality V$_2$-VI$_3$ topological insulators due to a smaller lattice mismatch \cite{Zeng2013}. InAs(111) in particular, has been under investigation as a material for novel n-channel metal-oxide-semiconductor field-effect transistors (nMOSFETs), where access to the L-valley conduction is believed to improve the technology \cite{Sumita2021}. Furthermore, epitaxial heterostructures of narrow-gap IV-VI and III-V semiconductors are a promising candidate for mid-infrared photonics, where the main benefits are the lattice constant matching, mechanical strength, chemical stability, and thermal conductivity as compared to the currently used substrates \cite{Haidet2021}. InAs is also the compound semiconductor of choice for building a hybrid semiconductor-superconductor based topological quantum bit, due to its unique combination of electronic properties such as e.g. large spin-orbit interaction and a narrow direct bandgap \cite{Thomas2018}. InAs(111) forms a low strain interface with GaSb and AlSb, other technologically relevant III-V compound semiconductors, as well as several II-VI compound semiconductors. Hence, a better understanding of the growth of InAs(111)A would open up a range of new device architectures. Herein, we focus on A-polar samples as hillock formation has been demonstrated to be suppressed compared to B-polar substrates. The wider triangular hillocks found on the B side generally lead to poorer optical and electrical properties.\cite{Yerino2017}
    
    The resurfaced interest for InAs(111)A has led to efforts of systematically mapping the MBE growth parameter space by Vallejo and coworkers in 2019 \cite{Vallejo2019}. Optimized MBE growth conditions are reported operating with As$_4$ species, based on reduction of root mean square (RMS) line roughness (Rq) values, extracted from large-scale (1$\times$1 $\mu\mathrm{m}^2$ and 5$\times$5 $\mu\mathrm{m}^2$) ambient AFM topographies and an increase in integrated photoluminescence intensities. With the constant miniaturization of devices, however, the atomic scale morphology and atomic impurities are becoming more important and reports on these are still rare. Commonly employed growth characterization tools such as ambient AFM, lack atomic resolution. Electron-based methods such as reflective high-energy electron diffraction (RHEED) or transmission electron microscopy (TEM) average over too large areas or volumes to yield a representative image of atomic impurities. STM on III-V materials on the other hand, provides the atomic resolution and additionally yields elemental contrast \cite{Timm2009}, allows detection of charge impurities \cite{Ebert2003} and probes the electronic properties with atomic lateral precision \cite{Suzuki2019, Batabyal2022}. By employing an MBE with an in-situ STM, alongside conventional characterization methods we examine the MBE growth parameter space for the InAs(111)A MBE homoepitaxy to find areas of optimized surface quality on the atomic- and large-scale.
    
    A series of samples with increasing As$_2$ flux, based around the previously found optimal conditions for As$_4$-based homoepitaxy \cite{Vallejo2019} reveals two scale-dependent regimes, by optimization of carefully extracted root-mean-square surface roughness (Sq) values. One regime on the small scale, as probed by STM, at an As$_2$/In flux ratio of 15 where the atomic scale surface only contains a single type of vacancy impurity alongside native charge impurities. We identified this impurity as an additional In-vacancy by comparison to density functional theory (DFT) based STM simulations and extract an approximate density of these of $4\times10^{11}$ cm$^{-2}$. Another regime on the large scale we expose an optimized pure step-flow regime at an As$_2$/In flux ratio of 26 by AFM imaging. Furthermore, we provide a method for employing the root-mean-square surface roughness Sq as a metric for surface quality on both scales and qualify it by comparison to technologically relevant metrics. Finally, we map these results to the growth parameter space to outline areas of interest for future investigations.

\section{\label{sec:experimental}Experimental}
\subsection{\label{sec:synthesis}Molecular Beam Epitaxy}
    Samples were grown and analyzed in an ultra-high vacuum (UHV) cluster tool, details of which can be found in the Supplemental Material \cite{Supp}. A series of six InAs(111)A samples was grown by homoepitaxial MBE. Four samples with increasing As$_2$/In flux ratio (3,15,26 and 33), keeping the indium flux constant at a corresponding growth rate of 0.07 ML/s and at a substrate temperature of T$_{\mathrm{sub}}\sim 500^\circ$C (Samples 1-4) and two at fixed As$_2$/In flux ratio with $\pm20^\circ$C offset from the initial sample temperature (Samples 5, 6). Each sample consisted of a 10$\times$10 mm$^2$ cleaved piece of single-side polished InAs(111)A wafers (Wafer Technology Ltd., n-doped, $\pm 0.1^\circ$ miscut). The pieces were gallium-glued to sample plates. After an initial ten hour degas in the main tunnel loadlock at 200$^{\circ}$C the samples were further degassed in a dedicated heater station at 250$^{\circ}$C for one hour. The surface oxide was removed by ten minutes of atomic hydrogen cleaning \cite{Khatiri2004} at the same temperature and with a hydrogen background pressure of $2.2\times10^{-5}$mbar. This procedure resulted in sharp RHEED patterns of the only stable 2$\times$2 surface reconstruction even before any buffer growth \cite{Veal2000}. Samples were kept at elevated temperatures (200$^{\circ}$C) until the background pressure had recovered to $\sim 1\times10^{-9}$mbar and were then transferred to the MBE chamber.

    Effective As$_2$ fluxes for As$_2$/In flux ratio calculations were calibrated using the GaAs(001)-(2×4)/(4×2) RHEED transition at a pyrometer sample temperature of 600$^{\circ}$C. Effective indium fluxes were calibrated by RHEED oscillations on InAs(001). Growth rates herein are given in monolayers (ML) of InAs(001) per second. The substrate temperature was ramped up with fully open arsenic-valve and measured using a pyrometer. The nominal thickness of the grown InAs(111)A buffer was 50 nm, with the exception of sample 2 which has a nominal thickness of 120 nm. The low indium determined growth rate of 0.07 ML/s was necessary to achieve the high As$_2$/In flux ratio and was kept constant for all experiments to avert rate-dependent effects.

    To observe the growth mode transition from islands growth to a step flow regime, RHEED intensity oscillations were recorded on a quarter of a similar 2-inch wafer for increasing arsenic flux. Due to suppressed island nucleation in a pure step-flow regime the vanishing of RHEED oscillations with increasing arsenic flux indicates the mentioned growth mode transition \cite{Neave1985}. The transition onset value of the As$_2$/In flux ratio was estimated by linear extrapolation of the maximum oscillation amplitude with increasing flux ratio to zero. The details of which are included in the Supplemental Material \cite{Supp}.

    It is generally known that the final ramp down can have an influence on the atomic scale cleanliness. Therefore, the procedure to ramp down the sample temperature after growth used in this study is inspired by the ramp for GaAs(001) homoepitaxy by LaBella et \textit{al.} where the goal was to achieve a well-ordered surface for STM. \cite{LaBella1999} The basic principle is to ramp the sample temperature as well as As-background down as soon as the growth is done. A similar procedure was shown to preserve the adatoms densities at growth conditions \cite{Kanisawa2013}. We chose to keep the As background pressure at growth conditions value and ramp the temperature from the value during growth to 350$^\circ$C at a rate of 2$^\circ$C/s. At this point the As valve and shutter were closed, immediately followed by ramping the substrate temperature to 250$^\circ$C at the same rate. 
    
\subsection{\label{sec:}Scanning Probe Microscopy}
     STM topographic images were recorded at liquid helium or nitrogen temperatures (4.3K or 78K), employing commercially available, electrochemically etched tungsten tips. Typical scan parameters for the topographic images consisted of a tip bias of V$_\mathrm{bias}=$0.5-2.0V and tunnel current set-point of I$_\mathrm{T}=$ 70-100 pA for typical scan sizes of $1\times1 \mu$m$^2$ to $40\times40$nm$^2$. The details of the STM (Scienta Omicron LT Nanoprobe 4P) can be found in the Supplemental Material \cite{Supp}. Impurity densities were extracted using the thresholding particle analysis tool in the MountainsSPIP{\texttrademark} software produced by Image Metrology A/S.

    An AFM (Dimension Icon, Bruker), operating in ambient conditions and using Bruker's exclusive PeakForce quantitative nanoscale mechanical characterization (QNM) mode was used to characterize the topography of samples after they were removed from the STM. Representative large scale overviews of the sample surface were acquired. Scan sizes of maximal 10$\times$10$\mu$m were taken on one to three locations (see Table in Supplemental Material \cite{Supp}) spread evenly over the sample surface or at representative locations as determined manually by optical microscopy. Terrace widths ($\mathrm{w}_\mathrm{Terrace}$) were extracted by manually measuring line the width of 10 consecutive terraces perpendicular to the step edges for at least 10 different local areas using MountainsSPIP{\texttrademark}.
    
\subsection{\label{sec:computational}Computational Methods}
    Density functional theory (DFT) calculations were performed using the Vienna Ab-Initio Simulation Package (VASP) code \cite{Joubert1999, Kresse1996, Kresse1996a, Kresse1993, Kresse1994} with the projector-augmented wave (PAW) method.\cite{blochl1994projector, Kresse1999} To describe the exchange-correlation interactions between electrons, the generalized gradient approximation (GGA) of Perdew, Burke, and Ernzerhof (PBE) \cite{perdew1996generalized} was used with a Hubbard U correction within Dudarev’s formalism \cite{dudarev1998electron} applied to the p-orbitals of both In and As.  $U_{\mathrm{eff}}$ values were determined using a Bayesian optimization method \cite{Yu2020}, and were found to be -0.5 eV and -7.5 eV for In and As, respectively. These values have been used successfully for various simulations of InAs surfaces and interfaces.\cite{yang2021principles, Yu2020, yu2021dependence} Spin orbit coupling was applied throughout \cite{steiner2016calculation} and a dipole correction was applied along the z-axis of the surface.\cite{neugebauer1992adsorbate} For all self consistent field (SCF) and density of states (DOS) calculations, $\Gamma$-centered Monkhorst-Pack k-point grids with densities of 5$\times$5$\times$1, and a cutoff energy of 350 eV were used.
    
    STM simulations were performed using the Tersoff and Hamann approximation, which states that the tunneling current is proportional to the local density of states (LDOS).\cite{Tersoff1985} To generate the pristine 2$\times$2 surface reconstruction, an In-terminated InAs(111) surface was constructed with 40{\AA} of vacuum, using an InAs lattice parameter of 6.0584{\AA}. Pseudo-hydrogen atoms with charges of 0.75 e\textsuperscript{-} were appended to the bottom surface to passivate the dangling bonds that result from cleaving of the surface \cite{Dardzinski2022}, and a single In atom was then removed from the top surface. Next, ionic relaxation of the top four atomic layers and the pseudo-hydrogen positions was performed until the Hellman-Feynman forces acting on ions were below 0.001 eV/{\AA}. During relaxation the As atoms in the second atomic layer moved to 0.09{\AA} below the top In atoms, putting them on approximately the same plane, correctly reproducing the expected surface reconstruction.\cite{bohr1985model}  After the relaxation finished, it was found that As dimers formed near the site of the defect. To model the In-vacancy, the relaxed InAs(111)A-(2$\times$2) surface was used as a starting point and an additional In atom was removed from the center of a 4$\times$4 surface unit cell. For the InAs(111)A-(2$\times$2) surface model and the In-vacancy surface 16 and 10 atomic layers were used, respectively.
    
    \begin{figure}[t] % Fig 1: atomic clean surface
    \includegraphics[width=0.5\linewidth]{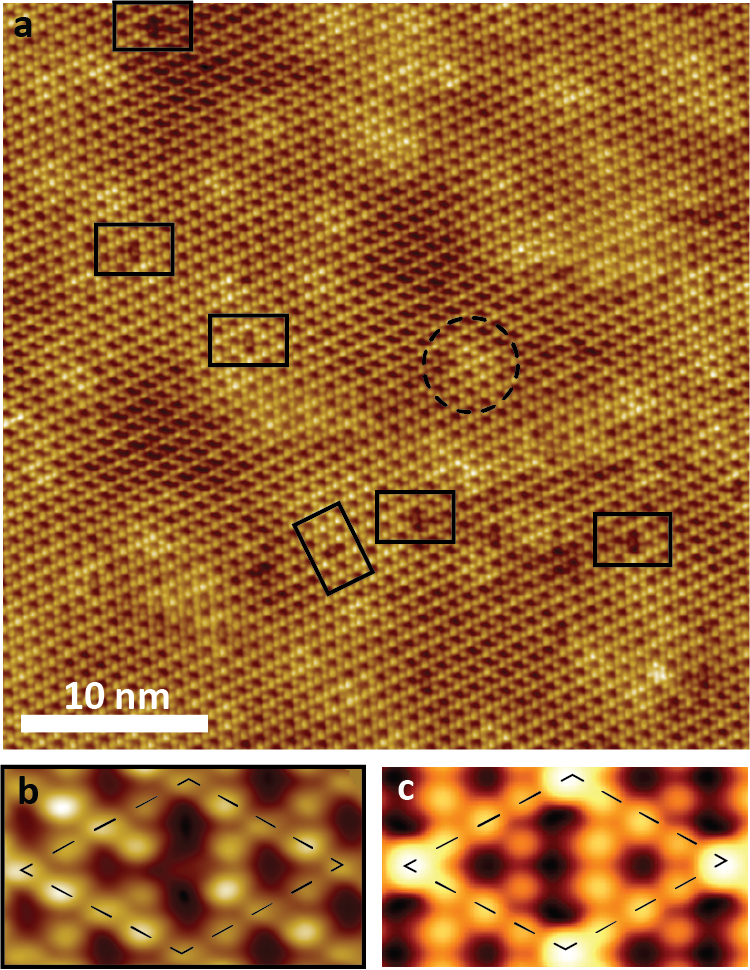}
    \caption{\label{fig:1}The indium-adatom deficient InAs(111)A surface with distinctive point defects. \textbf{a} Atomically resolved STM image (V$_\mathrm{bias}=400$ mV, I$_\mathrm{T}=100$ pA) revealing the InAs(111)A-(2$\times$2) reconstruction and six similar point defects marked by black rectangles. Faint bright protrusions (one exemplary marked by dashed circle) can often be interpreted as buried impurities. \textbf{b} Close-up of one of the observed point defects. The 4$\times$4 surface unit cell of this defect is indicated with dashed black lines. A comparison to a simulated STM image of an extra In-vacancy defect in the InAs(111)A-(2$\times$2) reconstruction in \textbf{c}, shows a good fit.}
    \end{figure}
    
    \begin{figure}[t] % Fig 2: III/V STM/AFM
    \includegraphics[width=\linewidth]{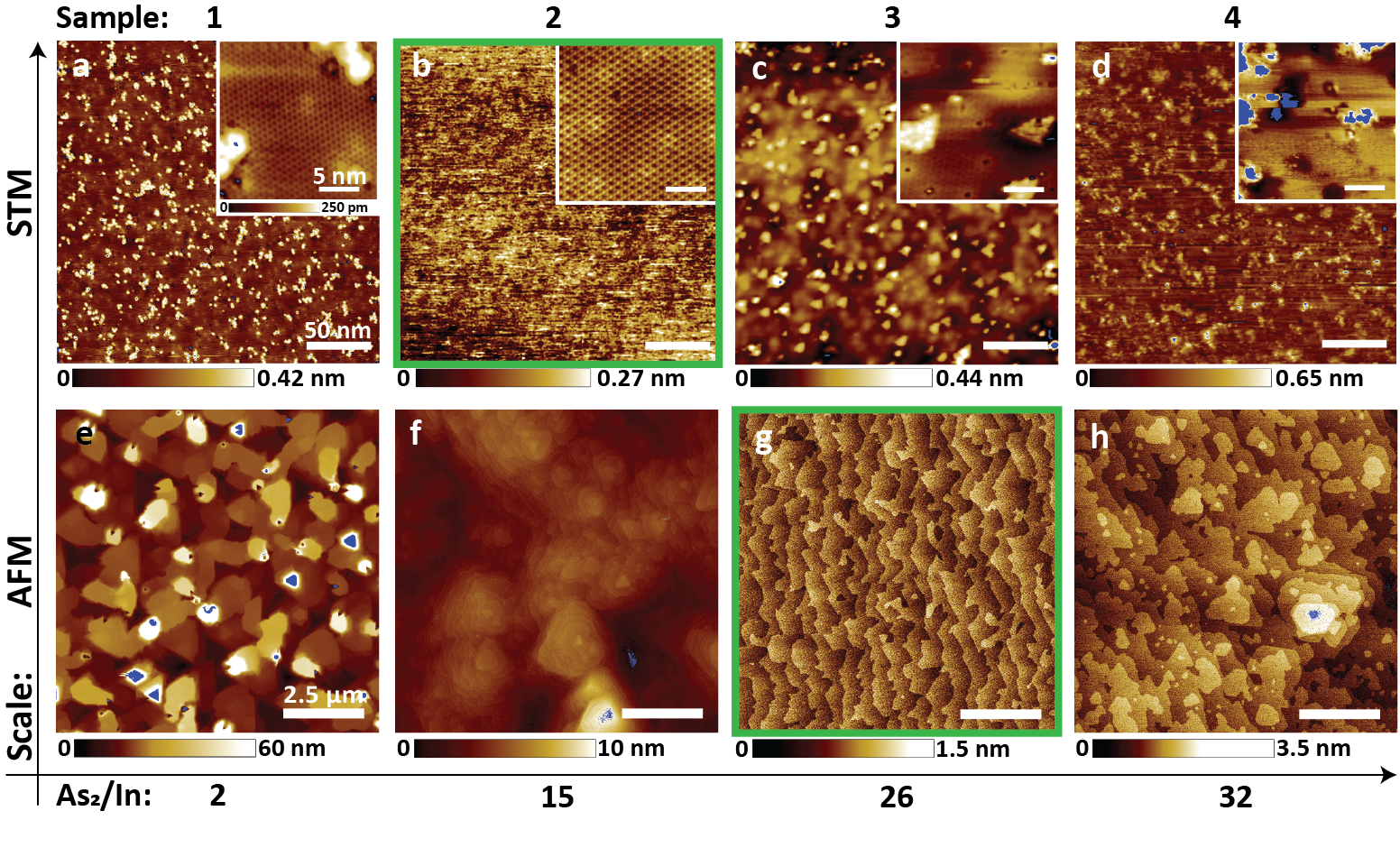}
    \caption{\label{fig:2} Comparison of STM to AFM of a series of growths with increasing As$_2$/In flux ratios at fixed sample temperature and indium flux. \textbf{a-d} STM topographic overview images (250$\times$250 nm$^2$) images of samples 1-4, respectively, purposely excluding step edges. Insets show a zoom-in (20$\times$20 nm$^2$) on the overview presenting atomic resolution and typical surface defects. \textbf{e-h} AFM topographic overview images (10$\times$10 $\mu$m$^2$) of samples 1-4, respectively. The highest quality surfaces for the respective scales are marked with green frames. Outliers are marked in blue throughout the figure and are either due to AFM scan artifacts (c, d, e) and due to a purposely imposed color scale limit to increase contrast here (a, f, h).}
    \end{figure}
    
\section{\label{sec:results_discussions}Results and Discussions}
    The combination of an MBE with in-situ transfer to an STM has enabled us to identify a set of MBE growth parameters for the InAs(111)A homoepitaxy, resulting in negligible amounts of indium adatoms and a low density of point defects as displayed in Figure \ref{fig:1}. Comparing the STM topographic image to DFT based simulated STM images we were able to identify the remaining impurity.  Figure \ref{fig:1}a shows a representative atomically resolved STM topographic image of the InAs(111)A surface where black rectangles mark the positions of the singular type of observed surface impurity. Additionally, local variations in the brightness (example marked with a dashed circle in Figure \ref{fig:1}a) hint at charge accumulation areas \cite{Ebert2003}, most likely caused by indium antisites native to the InAs(111)A homoepitaxial growth \cite{Kanisawa2008}. In Figure \ref{fig:1}b a close-up of the marked prevailing defect in the InAs(111)A-(2$\times$2) In-vacancy reconstruction is shown and the 4$\times$4 surface unit cell of this defect is marked with a black dashed rhombus. The comparison to a simulated STM image of an extra indium-vacancy impurity in the InAs(111)A-(2$\times$2) reconstruction in Figure \ref{fig:1}c shows a good agreement. We can, therefore, add this extra indium vacancy to the list of stable impurities of the MBE grown InAs(111)A surface, along with the indium and As$_2$ adsorbates \cite{Taguchi2006} and the previously mentioned indium antisites. Within the limited field-of-view of STM imaging, we calculated the density of these remaining impurities to be on the order of $4\times10^{11}$ cm$^{-2}$ by manually counting them in multiple selected areas of 40$\times$40 nm$^2$, spread over the sample. For comparison, in a recent publication by Chung and coworkers \cite{Chung2021} an impurity density in GaAs of $n\leq 1.5\times10^{11}\mathrm{cm}^{-2}$ was considered a low-density regime where defects are purely due to MBE source impurities. Post STM, ex situ AFM revealed that the larger scale topography of the atomically clean surface contains a large amount of hillocks. This could be the reason for the vanishing of native indium adatoms by preferential incorporation into the hillock sites of remaining indium adatoms during the ramp-down of sample temperature.\\

     To identify the optimal growth conditions for both atomically and large scale smooth surfaces the surfaces with increasing arsenic flux were probed first with in-situ STM followed by ex-situ AFM. For an overview of the respective imaging conditions and growth details, consult the table in the Supplemental Material \cite{Supp}. The As$_2$ flux was set to result in As$_2$/In flux ratios of approximately 2, 15, 26 and 32 while keeping the In flux and the substrate temperature roughly constant (samples 1-4). Representative STM and AFM images of each sample are show in in Figure \ref{fig:2}. The top row (Fig. \ref{fig:2}a-d) contains 250$\times$250 nm$^2$ STM topographic images and the insets show 20$\times$20 nm$^2$ zoom-ins with atomic resolution. Surface impurities change in amount and type with increasing As$_2$ flux (from left to right). In sample 1, adatom clusters and holes prevail while sample 2 is the optimised surface, as discussed previously in Figure \ref{fig:1}. Sample 3 contains both In- and As-adatoms and many small often triangular shaped islands. Finally, sample 4 contains In-adatoms and amorphous clusters of particles of unknown origin, most probably indium adatom clusters. The small scale impurities of sample 1 and 4 are comparable in type and amount. We notice that the different appearance of the surface reconstruction in the atomically resolved insets (hexagons in Figure \ref{fig:2}a, c and the "true" 2$\times$2 in Figure \ref{fig:2}b) is attributed to slight changes in STM tip shape and imaging quality which depends also partly on the surface quality. 

    Focusing on the classic AFM growth analysis in the bottom row of Figure \ref{fig:2}, we found a different set of optimal growth parameters, marked by a green frame, at which the amount of hillocks is greatly reduced. The representative AFM topographies shown have a size of 10$\times$10$\mu$m$^2$. 

\begin{figure}[t] % Fig 3: III/V metrics w/ histograms
    \includegraphics[width=0.5\linewidth]{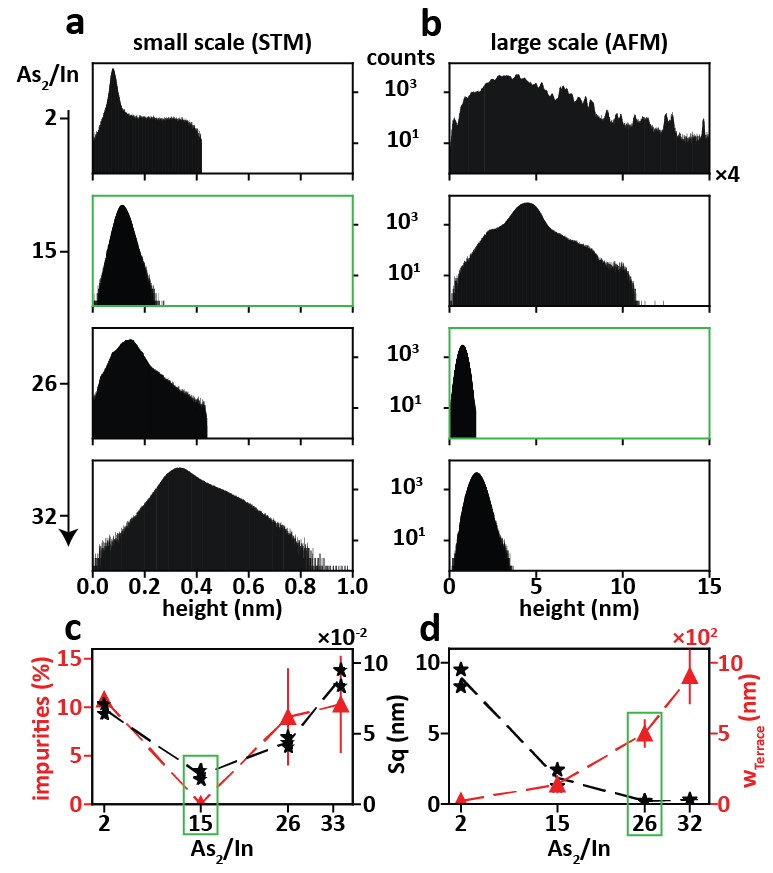}
    \caption{\label{fig:3}Height histograms of the least square plane leveled surfaces of the (\textbf{a}) STM and  (\textbf{b}) AFM topographies shown in Figures \ref{fig:2}\textbf{a-d} and \textbf{e-h}, respectively. Lower panel shows a comparison of the extracted RMS surface roughness Sq to technologically relevant metrics as a function of the As$_2$/In flux ratio for the respective scale. \textbf{c} STM extracted Sq compared to the STM-extracted impurity density in percentage of surface coverage by adatoms/islands. \textbf{d} AFM-extracted Sq values compared to the AFM extracted terrace widths. The respective dashed red and black lines are guides for the eyes.}
    \end{figure}

    A steady increase in terrace width from few tenths of nanometer to more than half a micrometer with increasing As$_2$ flux was extracted from manual measurements on zoomed-in versions and are visualized in the plot in Figure \ref{fig:3}d. The terrace width at optimized step flow conditions (Fig. \ref{fig:2}g) was found to be 500 nm. If we compare this to the average width of a terrace on the vicinal surface of an epi-ready InAs(111)A wafer the assumption that we grew in pure step flow is reasonable. Assuming formation of steps of one monolayer (ML) height $h = d_{\mathrm{InAs(111)}} = \frac{a_{\mathrm{InAs}}}{\sqrt{3}}$ with $a_{\mathrm{InAs}}=0.6058$ nm, following simple trigonometry, the typical miscut uncertainty of the used wafers $\Phi=\pm0.1^\circ$ will result at maximum in an average terrace width of $w = h_{\mathrm{step}}\cdot tan^{-1}(\Phi) = 200$ nm. Therefore, the miscut of the wafer of the sample grown in step-flow mode with 500 nm wide terraces (Fig. \ref{fig:2}a) would correspond to a miscut of $\Phi\approx0.04^\circ$, well within the margin of error.
    Regarding the formation of hillocks we observe an initial increase of hillock density with a peak at sample 2, followed by a reduced hillock density to levels on the order of initial larger scale defects on the bare wafer. This can also be observed easily by optical microscopy. Optical microscope images of both the bare wafer and samples 1-3 can be found in the Supplemental Material \cite{Supp}. It is noteworthy that this finding implies that by reducing the initial surface defects with e.g. a superlattice structure the appearance of hillocks in this regime might be able to be fully suppressed.

    The distribution of heights in least-square (LS) plane leveled surfaces is a good indicator for the surface quality. A narrow, normal-distributed heights histogram corresponds to a smooth surface. Histograms of the STM and AFM topographies in Figures \ref{fig:2}a-d and e-h, respectively are shown in Figure \ref{fig:3}a, b with the ordinates in a logarithmic scale to enhance the contributions from impurities (shoulders). The two optimized surfaces, marked by green boxes, show the narrowest distributions and lack significant shoulders when fitted by a normal distribution (not shown). To convert this visually encoded information into a single metric the surface roughness can be calculated.
    
     We now focus on the extraction method for the surface roughness and its use as a metric for surface quality. Previous reports by Vallejo et \textit{al.} relied on the metric of averaged root-mean-square (RMS) line roughness (Rq) as an indicator for high surface quality. We argue here, that RMS surface roughness (Sq) is less prone to being influenced by uni-directional defects such as the step edges at the atomic scale we are interested to quantify. The question remains, whether a carefully extracted Sq value can be used as a single metric to determine both atomic scale and larger scale surface quality as compared to the technologically relevant parameters of impurity density and terrace width, respectively. Sq would have the benefit of being easily deducible compared to the latter parameters. Therefore, in Figure \ref{fig:3}c (d) we plot the extracted Sq values from the STM (AFM) topographies in Figure \ref{fig:2}a-d (e-h) alongside the impurity density (extracted terrace width). 

    The small scale Sq values in Figure \ref{fig:3}c show a clear correlation with the surface density of impurities. For the large scale Sq values, we observe an anti-correlated behavior i.e. the surface roughness Sq decreases with increasing terrace width. 

    We ensured the relevance of the Sq values by excluding step edges. In case there is no terrace wide enough, masking of the other terraces was applied. Still, a direct inference of adatoms densities from the roughness values is not feasible due to multiple factors. Firstly, the quantity and quality of impurities has to be inferred from a topographic image and can not be captured in a single metric. And secondly, STM topographies always comprise a mix of topographic and electronic information. On the one hand this opens up a multitude of possibilities, on the other hand it makes it hard to deconvolute the two components. A solution can be to apply a high bias voltage, where electronic influences of the electronic structure on the topography are reduced. 

    The employed method to ensure the accuracy and relevance of the large scale Sq values consisted of a simple LS plane leveling. For very rough surfaces with many hillocks and wedding cake formation the observed anti-correlation may break down. In general, only the sum of multiple topographic images at different magnification will truly capture all the details of the surface quality in its complexity but the Sq values as extracted herein can aid as a fast, single-metric feedback for surface optimization.

    \begin{figure}[!ht] %Fig 4: Temp series
    \includegraphics[width=0.5\linewidth]{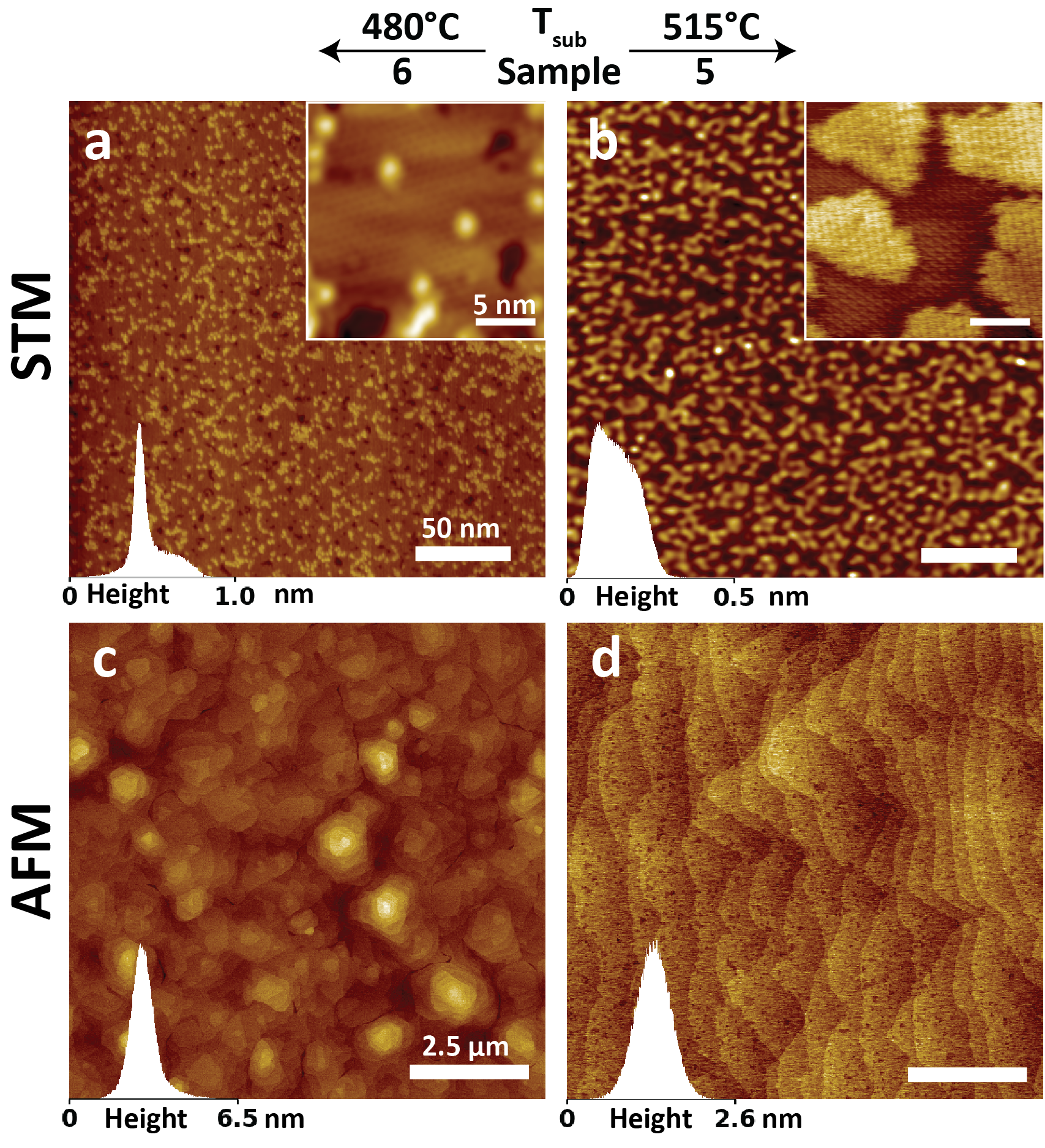}
    \caption{ \label{fig:4}Variations from the optimal substrate temperature of 500$^\circ$C at AFM optimized growth conditions. \textbf{a, b} STM topographic overview images (250$\times$250 nm$^2$), excluding step edges of samples 6 and 5 show an increase in surface coverage by impurities in both directions. Insets display atomically resolved close-ups in a region of the overview, respectively. \textbf{c, d} AFM topographies (10$\times$10 $\mu$m$^2$) reveal wedding cake formation at reduced substrate temperatures and step flow growth with smooth edges for an elevated temperature. Lower left insets show height distribution histograms of the respectively shown overview area on a logarithmic scale to enhance the low-counts shoulders.}
    \end{figure}

    Next, we investigate the temperature range around the previously found optimal substrate temperature of 500$^\circ$C to explore the extend of the discovered step-flow region horizontally in the growth parameter space. Figures \ref{fig:4}a, b (c, d) contain the STM (AFM) investigation at an increased and decreased substrate temperature, respectively for approximately constant As$_2$/In flux ratio values. Generally, we observe surfaces decrease in quality for these significant deviations of the optimal growth temperature of T$_\mathrm{sub}=500^\circ$C.
    On the atomic scale (Fig. \ref{fig:4}a, b), an increase in surface impurities for deviations in either direction from the optimal growth temperature are noted. At a lower temperature (Fig. \ref{fig:4}a) a mix of few nanometre small holes and clusters of adatoms are observed, similar to growth results over large ranges of growth parameters outside the optimal region, compare e.g. Ref. \cite{Vallejo2019}. At an increased T$_\mathrm{sub}=515^\circ$C (Fig. \ref{fig:4}b) the formation of a dense array of InAs islands of equivalent diameter of 5-10 nm are recorded. No such structures have been reported before and their appearance can likely be attributed to highly increased chances of decomposition of the In-As bonds above a temperature of $460-470^\circ\mathrm{C}$. Such conditions, in combination with the low growth rates employed, would then lead to an effective negative flux of indium atoms and the formation and growth of a network of vacancies which could appear like islands.  It may then be argued whether any growth is effectively achieved.
    
    At a reduced T$_\mathrm{sub}=480^\circ$C, AFM reveals that the step-flow regime is exited and formation of wedding cake structures commences (Fig. \ref{fig:4}c), indicating the lower boundary for step-flow in T$_\mathrm{sub}$ must be between 480-500$^\circ$C. The AFM topography at elevated T$_\mathrm{sub}=515^\circ$C (Fig. \ref{fig:4}d) still exhibits a step-flow like surface structure with unidirectional steps of reduced terrace width and smoother step edges as compared to T$_\mathrm{sub}=500^\circ$C. 

    \begin{figure}[t] %Fig 5: Map
    \includegraphics[width=0.5\linewidth]{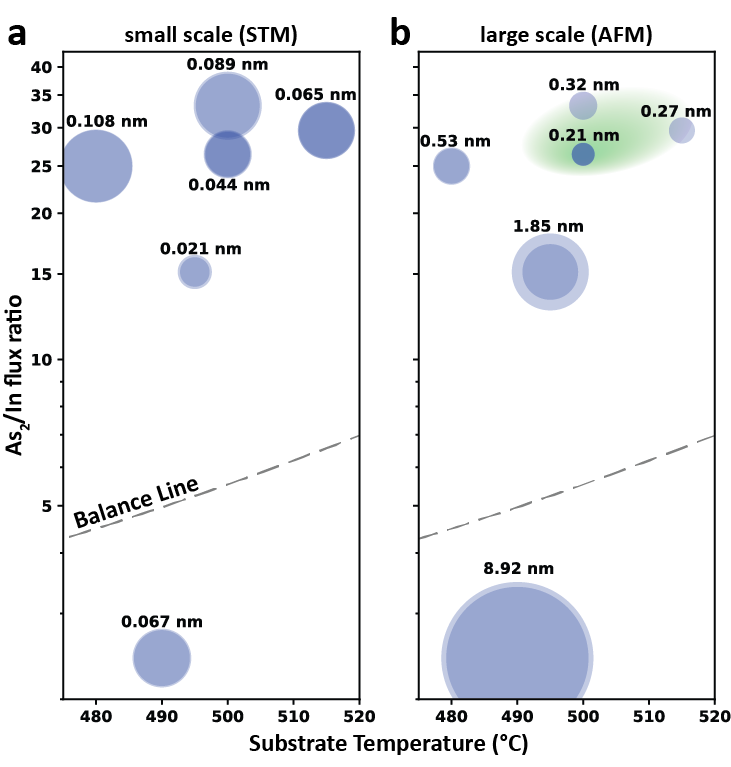}
    \caption{\label{fig:5} Mapping the (\textbf{a}) STM and (\textbf{b}) AFM Sq values in the growth parameter space reveals the two different regions of surface optimization for the two regimes. The balance line as established by Nomura et \textit{al.} \cite{Nomura1995} is shown as a gray dashed line. The location of it was confirmed for our setup, using the appearance of RHEED intensity decline and recovery (find details in the Supplemental Material \cite{Supp}). The green area in \textbf{b} indicates the approximate extend of the region with observed step-flow growth).}
    \end{figure}
    
    Finally, we map the extracted surface roughness values for all growths to a MBE growth parameter space to define regions of interest for future studies. This is shown in Figure \ref{fig:5}a,b for the small and large scale Sq values, respectively. The size of the circles corresponds to the surface roughness values and multiple circles (darker blue)) indicate multiple extracted values. The areas of smallest roughness (smallest circle diameter) have clearly different centers, depending on the scale one focuses on. RHEED studies by Nomura and coworkers who experimentally found the transition from arsenic flux limited growth to indium flux limited growth (often called detailed balance line) on InAs(111)A for MBE with As$_2$ species \cite{Nomura1995} are added to indicate the regions of indium and arsenic flux dependent growth rate regions. To validate the position of the detailed balance line for our setup we repeated similar RHEED measurements which can be found in the Supplemental Material \cite{Supp}. A good agreement between the balance line position we extracted and the Nomura et \textit{al.} data was found. Furthermore, the region where step-flow growth was observed is marked with a green shaded area in Figure \ref{fig:5}b. The sharper transition at a As$_2$/In flux ratio of about 25 is argued for by the vanishing of RHEED oscillations at this value (results see Supplemental Material \cite{Supp}), further indicating the onset of the step-flow region. It is, therefore, believed that no single set of growth parameters exists where the two optimized regions overlap.

\section{\label{sec:conclusions}Conclusions}
    In conclusion, we have demonstrated the special capabilities of an in-situ MBE-SPM combination as a feedback loop for atomic scale engineering of surfaces. We showed that at different scales surface optimization results in differing MBE growth parameters, implying the need for more atomic scale analysis during growth optimization for (111) substrate based nanoscale devices. The two regions of optimized large- and small-scale surfaces are achieved at substrate growth temperatures of 500$^\circ$C and 495$^\circ$C and As$_2$/In flux rations of 26 and 15. The former exhibits a pure step flow regime with unidirectional terraces of 500 nm and the latter up to 250 nm wide terraces with no indium adatoms and an extra indium vacancy as single prevailing surface impurity. We demonstrate the correlation of RMS surface roughness (Sq) values extracted in a careful manner with the technologically relevant metrics of terrace width and impurity density. The growth temperature space around the step flow yielding growth parameters was also probed and a dependence on the substrate temperature was observed with the surface structure changing noticeably for deviations of $\pm15-20^\circ$C. Finally, maps indicating the respective optimized regions in the MBE growth parameter space are provided as a base for focused engineering of a surface tailored to a specific application.

\section{\label{sec:contributions}Contributions}
    S.Z. prepared the samples, recorded most of the STM and all AFM topographies, analyzed the data and wrote the paper. R.B. contributed to study design, recorded some of the STM topographies and contributed to editing the paper. D.D. and N.M. contributed the DFT calculations and the editing of the related parts of the text. K.G.-R. and P.K. contributed to formulating the research question and editing of the paper.

\begin{acknowledgement}
    S.Z. acknowledges support from Microsoft Research, Microsoft Quantum initiative, Collaborative Project Description. Work at CMU was funded by the National Science Foundation (NSF) through grant OISE-1743717. The computer simulations used resources of the National Energy Research Scientific Computing Center (NERSC), a U.S. Department of Energy Office of Science User Facility operated under Contract No. DE-AC02-05CH11231.
\end{acknowledgement}
    
\bibliography{Main.bib}

% \documentclass[amsmath,amssymb,journal=cgd, manuscript=article]{achemso}
% \usepackage{graphicx}
% \usepackage{hyperref}
% \hypersetup{pdfencoding=auto,colorlinks=true,linkcolor=blue,citecolor=blue}

% \title[Scale-dependent optimised homoepitaxy of InAs(111)A]{Scale-dependent optimised homoepitaxy of InAs(111)A}

% \author{Steffen Zelzer}
% \affiliation{Center for Quantum Devices, Niels Bohr Institute, University of Copenhagen, 2100 Copenhagen, Denmark}

% \author{Rajib Batabyal}
% \affiliation{Center for Quantum Devices, Niels Bohr Institute, University of Copenhagen, 2100 Copenhagen, Denmark}

% \author{Derek Dardzinski}
% \affiliation{Carnegie Mellon University, Pittsburgh, PA 15213, United States}

% \author{Noa Marom}
%  \affiliation{Department of Materials Science and Engineering, Carnegie Mellon University, Pittsburgh, PA 15213, USA}
%  \affiliation{Department of Physics, Carnegie Mellon University, Pittsburgh, PA 15213, USA}
%  \affiliation{Department of Chemistry, Carnegie Mellon University, Pittsburgh, PA 15213, USA}

% \author{Kasper Grove-Rasmussen}
% \email{k\_grove@nbi.ku.dk}
% \affiliation{Center for Quantum Devices, Niels Bohr Institute, University of Copenhagen, 2100 Copenhagen, Denmark}

% \author{Peter Krogstrup}
% \affiliation{Center for Quantum Devices, Niels Bohr Institute, University of Copenhagen, 2100 Copenhagen, Denmark}

% \date{\today}
% \begin{document}
% \maketitle
    
\section{The UHV Cluster Tool}
\begin{figure}[!h]
    \includegraphics[width=0.5\linewidth]{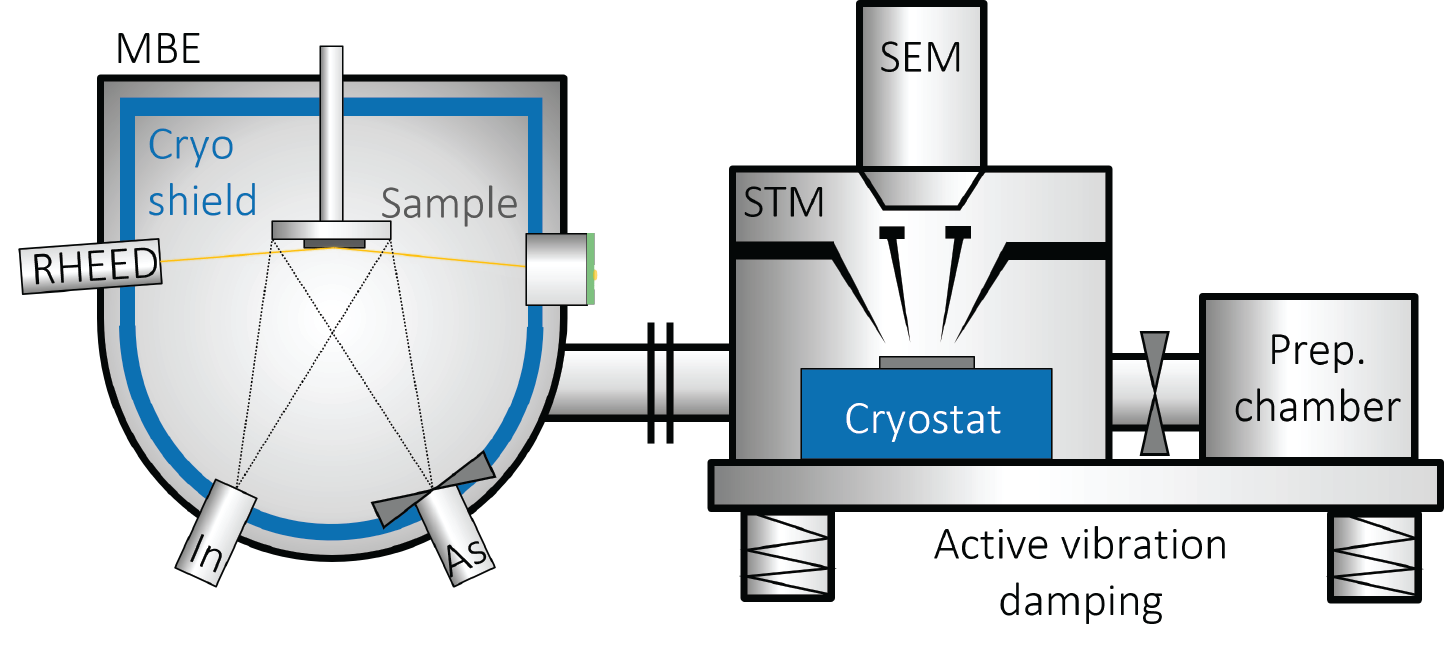}
    \caption{\label{fig:S_System} Schematic of the relevant parts of the MQML cluster tool for this study comprising of a MBE chamber, connected through a UHV tunnel to the STM system, which can be physically decoupled to reduce mechanical vibrations.}
    \end{figure}

    The used parts of the larger cluster consist of a MBE reactor with separate heater station, connected to the main tunnel (all Dr. Eberl MBE-Komponenten GmbH), and an STM (Omicron LT Nanoprobe 4P). Figure \ref{fig:S_System} shows a schematic drawing of the relevant parts of the cluster tool for this study. The MBE reactor is dedicated to planar growth of III-V semiconductors and amongst others equipped with an effusion cell for indium and a valved As source with cracker. Additionally, the RHEED gun and fluorescent screen are shown as well as the detachable UHV connection tunnel to the STM chamber.

\subsection{Samples}
    Table \ref{tab:samples} contains a summary of all samples considered in this paper along selected parameters and metrics of the STM and AFM measurements. 
    \begin{table}[!htb]
    \caption{\label{tab:samples} Overview of the samples and key MBE growth, STM (overview/close-up) and AFM parameters and metrics.}
    \begin{tabular}{ccc|cccc|cc}
                          \multicolumn{3}{c}{Growth}                      & \multicolumn{4}{c}{STM}                                                                 & \multicolumn{2}{c}{AFM}                                                        \\
    \hline
         \textbf{Sample} & \textbf{T$_\mathrm{sub}$} & \textbf{As$_2$/In}  & \textbf{250$\times$250 nm$^2$} & \textbf{T$_\mathrm{STM}$} & \textbf{V$_\mathrm{bias}$} & \textbf{I$_\mathrm{T}$}  & \textbf{10$\times$10 $\mu$m$^2$ } &  \textbf{No. of }\\
            \textbf{No.} & \textbf{[$^\circ$C]}      & \textbf{flux ratio} & \textbf{Sq [nm]}               & \textbf{[K]}              &  \textbf{[V]}              &  \textbf{[pA]}           & \textbf{Sq [nm]}                  &  \textbf{locations}\\
         \hline
        1 & 490 & 2.4  & 0.067 & 4.3 & 2.0/2.5 & 1000/100 & 8.92 & 2 \\ 
        2 & 495 & 15.0 & 0.021 & 300 & 1.0/0.4 &   80/100 & 1.85 & 2 \\ 
        3 & 500 & 26.4 & 0.044 & 78  & 0.5/-1.0&   20/50  & 0.21 & 3 \\ 
        4 & 500 & 32.3 & 0.127 & 78  & 1.0/1.0 &   70/50  & 0.32 & 1 \\ 

        5 & 515 & 29.6 & 0.065 & 78  & 0.9/0.7 &   70/200 & 0.27 & 1 \\ 
        6 & 480 & 25.0 & 0.108 & 4.3 & 2.0/2.5 &   100/70 & 0.53 & 2 \\ 
    \end{tabular}
    \end{table} 
    
\subsection{Sample transfer}
    Samples were transferred in-situ at pressures in the low $1\times10^{-9}$ mbar to mid $1\times10^{-10}$ mbar region to the STM chamber. Transfer took up to 20 min. No correlation of transfer times, pressure and atomic scale roughness or impurities was detected.
    
\subsection{LT Nanoprobe 4P STM}
    The STM is equipped with four independently operating piezo scanners and provides a scanning electron microscope (SEM) top view for basic SEM imaging and aligning the tips on nano-structures. One of the scanners can take up a qPlus sensor \cite{Giessibl2018} allowing atomically resolved non-contact atomic force microscopy (NC-AFM). Seamless switching between STM and NC-AFM allows to distinguish between topographic features and electronic information. Four additional internal electrical contacts allow for advanced device characterization methods being implemented in the future, such as e.g. scanning gate microscopy on devices during operation. The STM chamber can be physically separated from the UHV tunnel and rests on an active vibration damping platform. In addition, the STM stage is suspended from springs and Eddy-current damped to reduce mechanical vibrations to a minimum. 
    A bath cryostat allows operating temperatures of 4.3K, 77.4K or room-temperature depending on the used liquid gas. 
    In the adjacent preparation chamber, which also features the loadlock for externally loading samples and tips, typical STM reference samples can be prepared by Ar-sputter-anneal cycles. The chamber also offers a cold stage ($\geq$140K) with a custom cleaving mechanism to enable cross-sectional STM. A tip preparation tool (Ferrovac) enables deoxidation of in-house wet-etched tungsten tips and control of the tip apex radius via field-emission measurements.
    In-situ tip treatment consisted of voltage pulses up to $\pm$10V and gentle tip forming by driving the tip apex into the surface.
     
\section{RHEED}
\subsection{Step-Flow Transition}
    \begin{figure}[!h]
    \includegraphics[width=0.7\linewidth]{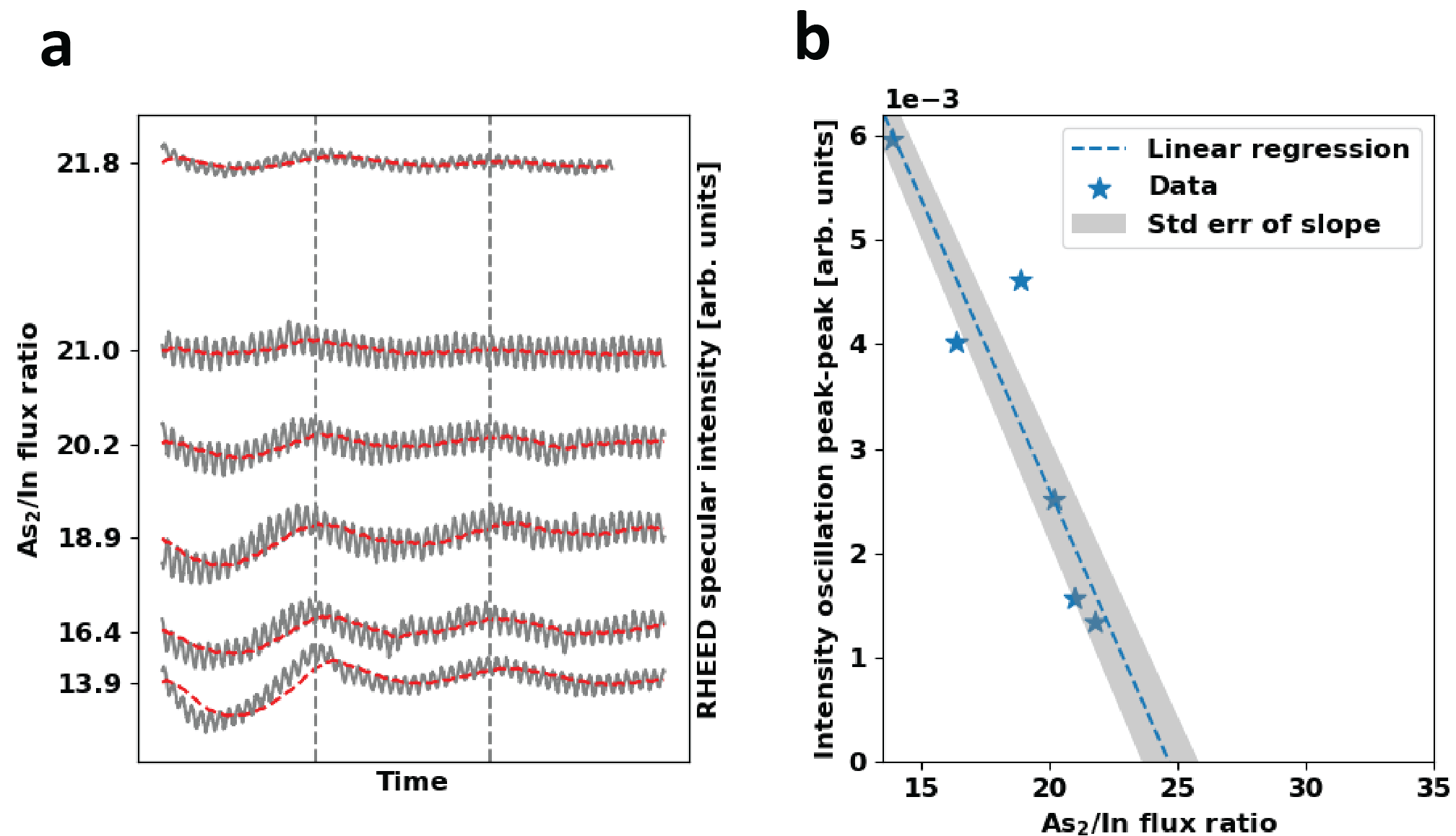}
    \caption{\label{fig:S1}\textbf{a} RHEED oscillations with increasing As$_2$/In flux ratio at a constant T$_\mathrm{sub} = 500^\circ$C. Grey lines are the raw intensity data and red dashed lines are the result of a first order Butterworth filter with a critical frequency of the order of the growth rate. \textbf{b} Plot of the maximal peak to peak oscillation amplitude over the As$_2$/In flux ratio. Extrapolation to zero oscillation amplitude indicates a transition to the step flow growth regime at a As$_2$/In flux ratio of 23-26 in accordance with the observed surface grown at a As$_2$/In flux ratio of 26.} 
    \end{figure}
    The growth parameter space that results in step-flow like growth on the other hand is comparably small, involves high As to In flux rations and is centered around about 500$^\circ$C substrate temperature. Due to a limit to the highest achievable As flux (especially in our system but generally for all valved cracker sources) the group III limited growth rate must be kept low (below 0.1 ML/s) to enable these high As$_2$/In flux ratios which is coincidentally to the benefit of the quality of the grown surfaces \cite{Vallejo2019} but might post a drawback for device fabrication. We experimentally probed the transition from step flow like growth to the onset of 2D nucleation by the appearance of RHEED oscillations upon ramping down the As flux at a fixed substrate temperature. The lack of a sharply defined transition was attributed to the low growth rates and subsequent low response time of the RHEED signal. At 0.06 ML/s formation of a single ML takes 16.6 seconds which is difficult to detect in the RHEED signal. Figure \ref{fig:S1} shows the results of such a RHEED oscillation series for the most relevant substrate temperature (T$_\mathrm{sub}$) of 500$^\circ$C. It can be observed that the maximal observed peak to peak oscillation amplitude in Figure \ref{fig:S1}a decreases with increasing As$_2$/In flux ratio. When plotted against each other as shown in Figure \ref{fig:S1}b and extended to zero we gain a range of As$_2$/In flux ratios of the transition region to step flow growth mode centered around the conditions for which we found the smallest roughness values.
    
\subsection{Balance Line Detection}
    \begin{figure}[!ht]
    \includegraphics[width=0.5\linewidth]{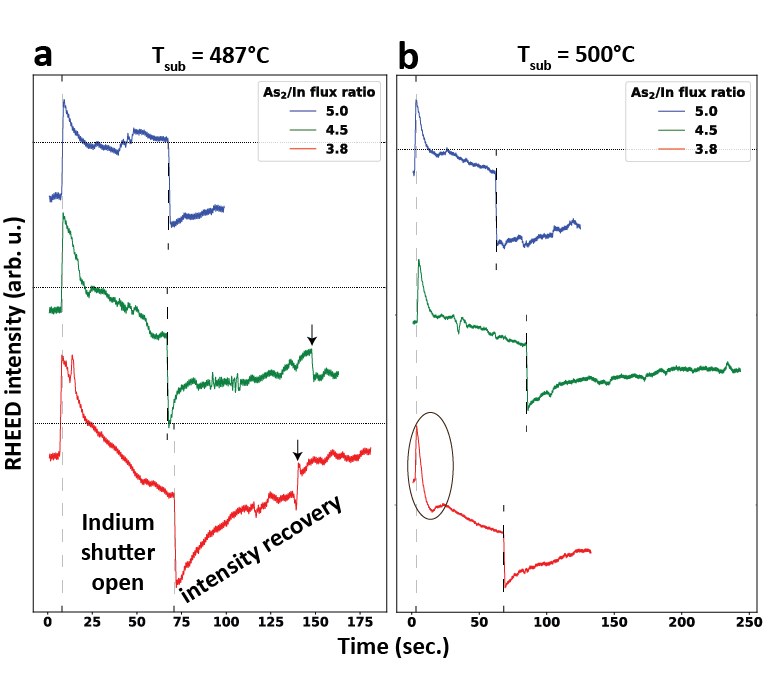}
    \caption{\label{fig:RHEED_int_recovery}RHEED intensity plotted over time for three different As$_2$/In flux ratios at \textbf{a} substrate temperature T$_\mathrm{sub}=487^\circ$C and \textbf{b} T$_\mathrm{sub}=500^\circ$C. The opening and closing of the indium cell shutter causes a sudden change in intensity (reflection). With decreasing As flux a more pronounced decrease in intensity during InAs growth is observed, indicating formation of a rougher surface, which also takes longer to recover.}
    \end{figure}
    The so-called detailed balance line for the InAs(111)A homoepitaxy is the border in the growth parameter space between group V and group III limited growth regimes. With increasing substrate temperature the arsenic will first start desorbing from the InAs(111)A surface from about 350$^\circ$C on and from about 520$^\circ$C on indium atoms will also start absorbing. Therefore, two regimes should be expected. An exponential increase in arsenic flux is needed from 350$^\circ$C to 520$^\circ$C to stabilize the surface, after which a stark increase would be expected to be needed to furthermore counter indium desorption, in other words, bury the surface to stabilize it. The balance line for InAs(111)A was recorded by RHEED intensity measurements by Nomura et \textit{al.} already in 1995 \cite{Nomura1995}. To confirm the validity of this data for the MBE in this study, we repeated such measurements for two points of the substrate temperature (478 and 500$^\circ$C) and three As$_2$/In flux ratio values, which should be crossing the balance line. We indeed found a change from almost constant RHEED intensity during a short growth period to a decrease in intensity as presented in Figure \ref{fig:RHEED_int_recovery}. Horizontal dotted lines show the level of RHEED intensity ignoring the typical initial growth related decrease in intensity, marked by a solid oval in Figure \ref{fig:RHEED_int_recovery}b after opening the In-shutter to grow approximately 5 ML of InAs(111)A at a rate of 0.076 ML/s, after which the In shutter is closed again and possible recovery of RHEED intensity is recorded. The opening and closing of the In shutter is visible in the RHEED signal as a sharp increase and decrease in intensity, respectively and is marked additionally with dashed vertical lines. Arrows in Figure \ref{fig:RHEED_int_recovery}a mark often observed shifting artifacts in the data that could not be assigned an origin. By these considerations, we estimate the balance line to be located between As$_2$/In flux ratios of 4.5 and 5 at T$_\mathrm{sub}=487^\circ$C and slightly above 5 for T$_\mathrm{sub}=500^\circ$C in good agreement with the Nomura data set where the respective values are 4.8 and 5.5.

\section{Optical Microscopy}
    \begin{figure}[!ht]
    \includegraphics[width=\linewidth]{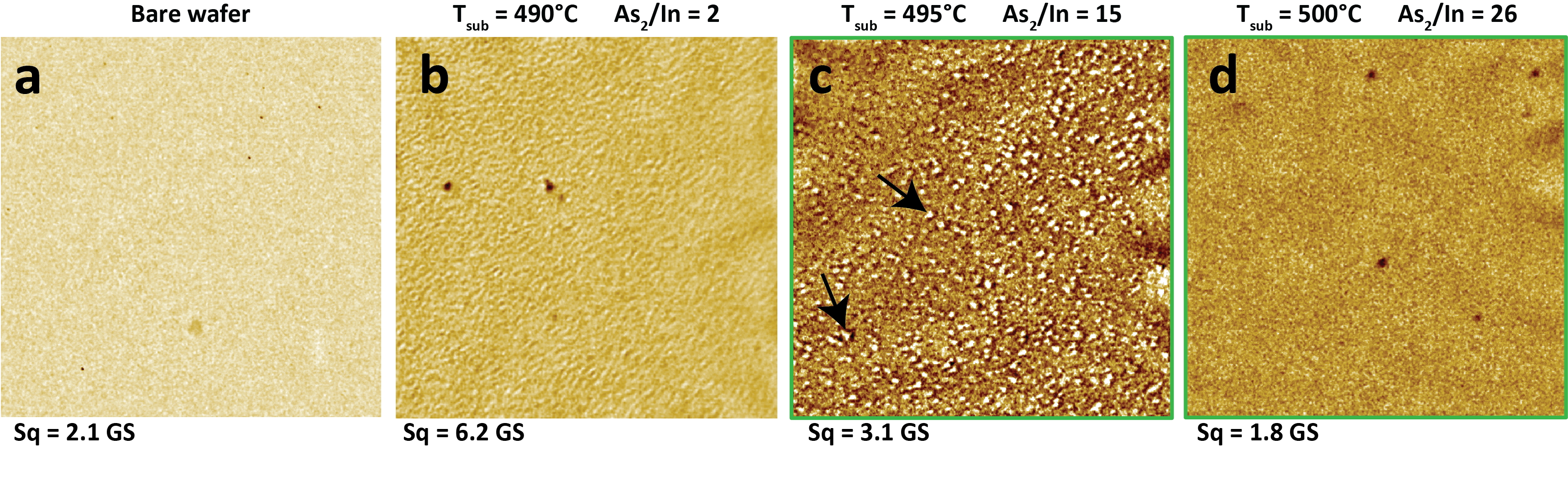}
    \caption{\label{fig:S3} Optical micrographs of \textbf{a} bare (epi-ready) InAs(111)A wafer \textbf{b-d} the first three samples in the As$_2$/In flux ratio series. Optimized surfaces marked by green frames. The arrows in (\textbf{c}) point out examples of hillocks.}
    \end{figure}
    In order to determine the growth quality at a much larger scale, optical microscope (OM) images were taken with the optical microscope which is aiding cantilever placement in the Bruker Dimension Icon AFM. They represent the respective areas where AFM scans where taken. Figure \ref{fig:S3}b-d contains OM images for the first three samples in the As$_2$/In flux ratio series. And Figure \ref{fig:S3}a an OM image of an epi-ready wafer surface. Grey-scale (GS) surface roughness values are extracted and show that the large scale optimal surface \ref{fig:S3}d is smoother than the bare epi-ready wafer. The wavy features are artifacts from the imaging of the AFM cantilever and of similar strength for all OM shown. Figure \ref{fig:S3}c has the highest amount of widely distributed sharply protruding hillocks, as marked by arrows, which could be the reason for the atomically clean surfaces in between, assuming adatoms preferentially are include in the hillocks.
    
% \bibliography{Main.bib}

% \end{document}
\end{document}